\documentstyle[12pt]{article}
\let\expandedversion=\iffalse
\let\useblackboard=\iftrue
\iftrue
\def\nbaselineskip{15pt}
\fi
\useblackboard
\font\blackboard=msbm10 scaled \magstep1
\font\blackboards=msbm7
\font\blackboardss=msbm5
\newfam\black
\textfont\black=\blackboard
\scriptfont\black=\blackboards
\scriptscriptfont\black=\blackboardss
\def\Bbb#1{{\fam\black\relax#1}}
\else
\def\Bbb{\bf}
\fi

\def\sect#1{\subsection{#1}\setcounter{equation}{0}}
\catcode`@=11
\def\@cite#1#2{\if@tempswa [#1]\else$^{\scriptscriptstyle
\mbox{\rm\scriptsize#1}}$\fi}
\newcommand{\eqn}{\begin{eqnarray}}
\newcommand{\enq}{\end{eqnarray}}
\newcommand{\eqa}{\begin{array}}
\newcommand{\ena}{\end{array}}
\newcommand{\eq}{\begin{equation}}
\newcommand{\en}{\end{equation}}

\newcommand{\no}{\nonumber}

\def\comments#1{}
\newcommand{\IZ}{{\Bbb{Z}}}

\def\1N{1\over N}
\def\CC{\Bbb{C}}
\def\CP{$\Bbb{C}P^n$}
\def\sCP{$\bf CP^n$}

\def\cp{\CP}
\def\C{$\Bbb{C}P$}
\def\RR{\Bbb{R}}
\def\ZZ{\Bbb{Z}}
\def\del{\partial}
\def\half{{1\over 2}}
\def\Tr{{\rm Tr\ }}
\def\im{{\rm Im\hskip0.1em}}
\def\bra#1{{\langle}#1|}
\def\ket#1{|#1\rangle}
\def\vev#1{\langle{#1}\rangle}
\def\CT{\cal T}
\def\Dslash{\rlap{\hskip0.2em/}D}

\begin{document}
\setlength{\unitlength}{0.25cm}
\begin{titlepage}
\hfill{\vbox{\hbox{{\sc OUTP-93}-24P}\hbox{RU-93-56\hfil}}}
\vspace{1.5cm}

\begin{center}
{\LARGE
Topological-Antitopological Fusion\\[0.1in]
and the Large $N$ $CP^N$ Model
}\\[1.5cm]

{
M. Bourdeau
\footnote{{\tt bourdeau@thphys.ox.ac.uk}}}\\
Dept. of Theoretical Physics\\
Oxford University\\
Oxford, OX1 3NP\\[0.5cm]
and\\[0.5cm]
{
M. R. Douglas
\footnote{{\tt  mrd@physics.rutgers.edu}}}\\
Dept. of Physics and Astronomy\\
Rutgers University\\
Piscataway, NJ 08855\\[1.5cm]

\end{center}

\vfil
\begin{center}
{\sc Abstract}
\end{center}

\begin{quotation}
We discuss the large $N$ limit of the supersymmetric $CP^N$ models
as an
illustration of Cecotti and Vafa's $tt^*$ formalism.  In this limit the
`$tt^*$ equation' becomes the long wavelength limit of the $2D$ Toda
lattice,
an equation first studied in the context of self-dual gravity.  We
show how simple finite temperature and large $N$ techniques determine the
relevant solution, and verify analytically that it solves the $tt^*$
equation, using Legendre transform techniques from self-dual
gravity.
\expandedversion (Expanded version.)
\fi
\end{quotation}
\vfill
December 1993 \hfill

\end{titlepage}
\setlength{\baselineskip}{\nbaselineskip}
\newpage

\sect{Introduction~}

Recently there has been a lot of work on $N=2$ supersymmetric
models in two dimensions, in the context of string theory and
integrable and massive quantum field theories. Non-renormalization
theorems for these models make them a lot easier to study
and classify.  Much structure is encoded in a finite dimensional closed
subalgebra of chiral primary fields.
Many models admit a Landau-Ginsburg description and these can be
largely understood in terms of their superpotential.
More recently, Cecotti et.~al.\cite{cfiv} offer a further
classification of models by defining a `new index'
$\Tr (-1)^F F e^{-\beta H}$ which exhibits even more of
the structure of the model. The new index depends only on $F$-term
perturbations, thus is much simpler than quantities like the free energy,
but unlike Witten's index $\Tr (-1)^F e^{-\beta H}$
and the chiral ring
it can encode information about scale and coupling dependence of the
model,
both at short distances (e.g. dimensions of perturbations near the UV
fixed point)
and at long distances (e.g. the soliton spectrum).

In principle the new index is exactly calculable in any two-dimensional
$N=2$ theory, whether or not it is integrable.
One first considers the inner product on the space
of supersymmetric ground states, which geometrically plays the
role of a metric.
This metric satisfies a differential equation as a function of the
couplings,\cite{cvtop}
essentially the same as the one found for Zamolodchikov's metric in the
case of
$N=2$ superconformal theory.\cite{special}
In many interesting cases it
reduces to a familiar equation of mathematical physics.  It is a short
step
from the metric to the new index (whose physical interpretation is perhaps
clearer).

In \cite[x]{cfiv}, the authors investigate the new index for several
integrable models, such as the $N=2$ sine-Gordon and minimal $N=2$
theories.
They show how to obtain the new index for integrable theories,
given the exact $S$-matrix, by means of the
thermodynamical Bethe ansatz.
This method requires solving a moderately tractable set of coupled
non-linear integral equations.
In \cite[x,y]{fs,fsz}, applications to polymer physics
(self-avoiding random walks) are carried out.
In all of these papers, the
authors uncover previously unknown mathematical structure of $N=2$
theories and equivalences between solutions of integral equations and
differential equations.
In the simplest case (the $A_2$ deformed minimal model), the
differential equation is a special case of Painlev\'e
 ${\rm I\!I\!I}$ (or the sinh-Gordon equation) and the relevant solution
was
shown numerically to be equal to the TBA result.

In \cite[x]{cvsig}, Cecotti and Vafa study supersymmetric $\sigma$ models
and
obtain a differential equation for the
metrics of the SUSY \C$^1$ and \C$^2$  models. This equation
is also the sinh-Gordon equation but the metric differs from the $A_2$
minimal
case by its different boundary conditions. For \cp\ with $n \ge 3$,
the equations have not been studied explicitly.

Historically, the \cp\ models have been studied more extensively than
the other models one might consider as applications,
and they are very interesting theories with some analogies to QCD:
they are asymptotically free, and they have instantons and
$\theta$ vacua, leading to a `$U(1)$ problem'
with a resolution like that of QCD.
(The supersymmetric case, which has fermions of zero bare mass, is more
similar to QCD with a massless quark.)
Moreover, the \cp\ model is very simple to solve in the large $n$ limit:
to leading order in $1/n$, $S$-matrix elements are given by summing tree
diagrams, while bulk quantities like the free energy are calculable by
simply
extremizing an effective action.

This leads to the question of whether the $tt^*$ methods show comparable
simplifications in this limit.
In this paper we study the large $n$ supersymmetric \cp\ model,
and find that the $tt^*$ equation determining the metric is
an equation first studied in the context of self-dual gravity, and
related (by a Legendre transform) to a symmetry reduction of
Pleba\'nski's `heavenly' equation for a self-dual K\"ahler potential
in $D=4$.
Few explicit solutions of this equation are known.

A calculation of the index (and metric) using large $n$ techniques
proceeds in
two steps.  One can write the model in terms of free fields parametrizing
$\CC^{n+1}$ with auxiliary fields implementing the reduction to \cp.
Integrating out the free fields gives a quantum effective action, and in
the
large $n$ limit observables (such as the index) are dominated by a saddle
point
of this effective action.
The second step is to minimize the effective action with respect to the
auxiliary fields.
It will emerge that in our problem, the minimization can be reinterpreted
as
precisely a combination of known techniques for finding solutions to
self-dual
gravity from those for simpler equations via Legendre transform.
Thus we will prove that the index and metric, computed independently,
solve the
$tt^*$ equation.

Our original motivation for this work was simply to have a field theory
example in which we could make every element of the $tt^*$ formalism
completely
explicit.
Perhaps the most useful consequence is that in this model, extensions to
the
original ideas, such as understanding the role of the higher couplings, or
of
changes to the two-dimensional space-time metric, can be studied
explicitly.
It may also be possible to study more interesting large $n$ models such as
Grassmannian target spaces or other models with $n^2$ degrees of freedom.

In section 2 we review the work of Cecotti, Vafa and collaborators
and discuss the $tt^*$ equations and new index.
In section 3 we review the supersymmetric \CP\ model,
and its solution in the large $n$ limit.
In section 4 we derive the large $n$ limit of the
$tt^*$ equation for the \CP\ model.
In section 5 we will derive its
solution and discuss the connections with self-dual gravity.
Section 6 contains conclusions.

\sect{The ground-state metric for $N=2$ theories and the new index}

The situation governed by the $tt^*$ equations is the following.
We have a $d=2$, $N=2$ supersymmetric field theory quantized on a
Euclidean manifold, with metric and boundary conditions preserving $N=2$
global
supersymmetry.
We assume there are a discrete set of supersymmetric ground states, that
at
least one dimension is compact, and that a Hamiltonian defined on a
compact
hypersurface has a gap.
Given all this, certain cleverly chosen correlation functions can be
reduced to
sums over the ground states.

For superconformal theory this is very easy to arrange; we just need to
have a
compact dimension with Ramond boundary conditions.  Since Weyl
transformations
act so simply the form of the metric is not important.
For general theories we expect some constraints.  First, to have an
unbroken global supersymmetry, there should be a covariantly constant
spinor on
the surface, which in two dimensions implies the metric is flat.
Second, for a correlation function to reduce to a sum over ground states,
it
must be expressible as a limit in which the distance between any pair of
operators inserted goes to infinity.  Thus the space-time must also have a
non-compact dimension.  It is natural to require this dimension to be
infinite
in both directions, in which case space-time is a cylinder.

In some sense the formalism is a particular case of the ``nonabelian
Berry's
phase'' we would see if we varied the parameters of any quantum system
with a
degenerate ground state.
The first role of supersymmetry in the discussion is simply to guarantee
that
there will be a set of exactly degenerate ground states.
$N=2$ supersymmetry came in when we identified these with deformations of
the
couplings of the theory; in general there is no relation but in $N=2$
spectral
flow provides this relation.

An $N=2$ theory can be `topologically twisted';\cite{ey,witten} one
component of the supercharge is singled out to play a role like the BRST
charge, and the state space projected to its cohomology.
The stress-tensor is also modified to be a BRST commutator,
so correlation functions
are independent of the positions of operators.
This truncation keeps only the ground states and makes the identification
of
these with operators very simple.  One way to describe the $tt^*$ results
is
that they answer the question: from the data of an $N=2$ theory that
survive
topological twisting, what can one reconstruct about the original theory?
However one does not need to bring the topological theory into the
discussion,
and as a rule we will not.

Let us then consider an $N=2$ theory on a cylinder, where the long
dimension
has coordinate $x^1$ and length $L$ (which will go to
infinity), and the compact dimension has coordinate $x^0$ and
circumference
$\beta$.  It is useful to have operator formalisms both with states
defined at
constant $x^1$ and with states on constant $x^0$.
In two dimensions the $N=2$ supersymmetry algebra
takes the form
\eqn\label{algebra}
Q^{+2}_L=Q^{-2}_L={Q}^{+2}_R&={Q}^{-2}_R=&\{Q^+_L,{Q}^-_R\}=
\{Q^-_L,{Q}^+_R\}=0\no\\
\{Q^+_L,{Q}^+_R\}=2\Delta&  & \{Q^-_L,{Q}^-_R\}=2\Delta^*\no \\
\{Q^+_L,Q^-_L\}=H-P &  & \{Q^+_R,Q^-_R\}=H+P \no\\
{}[F, Q^{\pm}_L]=\pm Q^{\pm}_L & &  [F,{Q}^{\pm}_R]=\mp{Q}^{\pm}_R \no\\
Q^{+\dagger}_L=Q^-_L & &  ({Q}^+_R)^\dagger = {Q_R}^-\\
Q^{\pm}\equiv{1\over\sqrt{2}}(Q^{\pm}_L+{Q}^{\pm}_R)& &\{Q^-,Q^+\} = H.
\enq
$F$ is called `fermion number,' and
$Q^{\pm}_L$  and ${Q}^{\pm}_R$ are left and right supercharges.
In non-compact space one can have a non-zero central term
$\Delta$, which will come in below.
For now we consider $x^1$ as time.

The most basic elements particular to a given $N=2$ theory are the chiral
and anti-chiral rings.
We list the chiral operators
$\phi_i$ satisfying $[Q^+,\phi_i]=0$,
and anti-chiral operators $\bar{\phi_i}$ satisfying
$[Q^-,\bar{\phi_i}]=0$.
The chiral ring is defined in terms of the operator product algebra as
\eq\label{ring}
    \phi_i\phi_j  =  \sum_k C^k_{ij}\phi_k + [Q^+,\Lambda].
\en
Since the derivative of any operator (and the stress tensor itself) is a
descendant under $Q^+$ (and $Q^-$), the positions of the operators on the
left
hand side do not matter.
The anti-chiral ring will have structure constants $\bar C^k_{ij}$.

Equally important are the supersymmetric (Ramond) ground states
\eq\label{susygr}
H|a\rangle = Q^{\pm}|a\rangle = 0.
\en
We could make a correspondence between these and chiral fields by choosing
a
canonical ground state $\ket{0}$.
Then we can identify
\eq
               \phi_i|0\rangle=|i\rangle+Q^{+}|\Lambda\rangle .
\en
Finally we could project on the true ground state by applying an operator
like
$\lim_{T\rightarrow\infty} \exp -HT$.
We could also do this with anti-chiral fields
$\bar{\phi}_i$, producing states to be called $|\bar{i}\rangle$.
The structure constants $C^k_{ij}$ then also give the action of the chiral
operators on the ground states:
\eq\label{chiralact}
  \phi_i|j\rangle  =  C^k_{ij}|k\rangle + Q^+|\psi\rangle.
\en

This construction is not completely satisfactory because it is not clear
that
the correspondence is one to one; furthermore it depended on the choice of
$\ket{0}$.
Both problems are dealt with by making a correspondence using
spectral flow.\cite{cvtop}
In principle, this constructs the state $\ket{i}$ by doing a
path integral on a hemisphere with an insertion of $\phi_i$.
We need spectral flow to put this state in the Ramond sector, and we can
think
of it as turning on a $U(1)$ gauge field coupled to the fermion number
current,
with holonomy $e^{i\pi}$ on the boundary.
We can then take as $\ket{0}$ the state produced by inserting the identity
operator $\phi_0\equiv 1$, and non-degeneracy of the two-point function
$\vev{\bar\phi_i \phi_j}$ will imply that the correspondence is one to
one.

Now we take $|i\rangle$ and $|\bar{j}\rangle$ to denote the
basis of
ground states corresponding to the fields $\phi_i$ and $\phi_{\bar{j}}$.
CPT will relate $|i\rangle$ to a state $\bra{\bar i}$ so the usual Hilbert
space metric will be the hermitian
\eq\label{metricg}
g_{i\bar{j}}=\langle\bar{j}|i\rangle.
\en
Another structure present in the theory is the ``real structure'' $M$
expressing one basis in terms of the other:
\eq\label{metricM}
    \langle\bar{i}|=\langle j|M^j_{\bar{i}}
\en
CPT implies $MM^*=1$.

A combination of these produces the `topological' metric $\eta$:
\eq
     g_{i\bar{k}}=\eta_{ij}M^j_{\bar{k}}
\en
This is the two-point function in the topologically twisted theory
and as such it is in many ways a more basic object than $g$ or $M$.
We will not use it in the following but instead refer the reader to the
extensive literature on topological field theory.

Supersymmetry-preserving perturbations of the action are of two types.
In general we need to write a commutator with all four supercharges
(or integral $d^4\theta$) to preserve all supersymmetries.
However we can also write
\eq\label{pertS}
     \delta S =\sum_i
\int d^2x~~ \delta t_i\{Q^-_R,[Q^-_L,\phi_i]\}
+\delta \bar{t}_i \{Q^+_R,[Q^+_L,\bar\phi_i]\}.
\en
where $\phi_i$ and $\bar\phi_i$ are chiral and anti-chiral fields.
A perturbation which can only be written in this form is called an
$F$ term; the others are $D$ terms.

Given a space of theories $T\in\CT$ defined by perturbing
around a base theory $T_0$ (and thus with coordinates $t_i$ and $\bar
t_i$),
we would like to define the rings, ground states and metric for each
theory
$T$.
A simple way to do this locally is to use
the same operator basis for the chiral ring for each $T$, but evaluate the
o.p.e. in (\ref{ring}) and the path integral in the spectral flow
construction
using the action for theory $T$.
This will give structure constants and ground states depending on the
couplings, and in principle from this we could compute the metric $g$ as a
function of the couplings.
There is an important subtlety in this computation.
One might think that since an operator
$\{Q^-_R,[Q^-_L,\phi_i]\}$ annihilates a supersymmetric ground state,
inserting
(\ref{pertS}) into (\ref{metricg}) would give zero.  It is true for $D$
terms,
but not for $F$ terms.
One sees the subtlety most simply by considering a mixed second derivative
of
the metric, which is evidently expressed as
\eq
\partial_k\bar\partial_l\vev{\bar j|i} =
\vev{\bar j| ~\int d^2x d^2x'~ \{Q^-_R,[Q^-_L,\phi_k(x)]\}
\{Q^+_R,[Q^+_L,\bar\phi_l(x')]\}~|i}.
\en
We do not have a perturbation of the states like (\ref{chiralact}),
followed by
projection on the true ground states, but rather some mixture of the two.
By considering the action of the supercharges in this formula, one sees
that it
differs by including contact terms where $x=x'$ (as in \cite[x]{kutasov}).

A nice way to disentangle these is to separate the dependence on the
couplings
of the states $\bra{\bar j}$ and $\ket{i}$, varying the action for the two
path
integrals we use to construct the two states.\cite{cvtop}
In the spirit of the non-abelian Berry's phase, define the gauge
connection
\eq
   A^{\hphantom{i}j}_{i\hphantom{j}k}=
g^{j\bar j'}\langle\bar{j'}|\del_i|k\rangle
\en
and its conjugate.
By definition, the metric $g$ is covariantly constant with respect to the
derivatives
\eqn
   D_i=\del_i-A_i& &\bar{D}_i=\bar{\del}_{\bar i}-\bar{A}_{\bar i}.
\enq
Then, we might expect covariant combinations like the curvature to be
especially simple.
Writing these out explicitly and manipulating the supercharges gives
\eq\label{holo}
 [D_i,D_j]  =  [ \bar{D}_{\bar i},\bar{D}_{\bar j}]=0
\en
and for the mixed case terms involving insertions of the Hamiltonian,
which can
be written as total $x^1$ and ${x'^1}$ derivatives.  Considering the
boundary
terms in the $x$ integrals, one limit will produce the same contact term,
which cancels in the commutator, while for the other limit, with all
operators
at large distances, the correlator reduces to a sum over ground states,
which is evaluated using (\ref{chiralact}).
Thus one finds
\eq
 [D_i,\bar{D}_{\bar j}]  = -\beta^2[C_i,\bar{C}_{\bar j}],
\en
a differential equation for the metric.
By (\ref{holo}) one can choose a basis in which
$\bar{A}_i=0$, so $A_i=g^{-1}\partial_ig$, and it becomes
\eq
  \bar{\partial}_{\bar j}(g\partial_ig^{-1})
  =\beta^2[C_i,gC^{\dagger}_{\bar j}g^{-1}]
\en

These are the $tt^*$ equations, which given enough boundary conditions
determine the metric $g$.
Different models with the same chiral ring can have different metrics:
thus the boundary conditions are a crucial part of the story.
In the cases considered in detail,\cite{cvtop,cvsig,cfiv}
the dependence on one relevant coupling is studied.  Let this define a
mass
scale $m$; then small $\beta m$ is weak coupling and this limit of the
metric
can be found using semiclassical techniques.
The large $\beta m$ boundary conditions are even simpler and are best
explained
in terms of the `new index' (see below).

It is perhaps worth noting that quantum field theory was not really used
in
deriving the $tt^*$ equations
(the original derivation of \cite[x]{cecotti} was in the
context of $N=2$ supersymmetric quantum mechanics!) and whatever quantum
field
theory structure is there is in some sense fed in through the boundary
conditions and the chiral ring structure constants.
It is a pleasant surprise then to find that quite non-trivial quantum
field
theoretic information emerges.
This point also holds out some hope that these ideas will have value in
$D>2$.
We recall as well that no assumption about
the integrability of the theory
(in the usual senses of having extra conserved charges
or a factorized $S$-matrix) was made.

\medskip
Another observable depending only on $F$ couplings was given in
\cite[x]{cfiv}.  Although it is simply related to the metric it has a
clearer
physical interpretation.
It is modeled after the index ${\rm Tr} (-1)^F e^{-\beta H}$,
which 
is completely
independent of finite perturbations of the theory for $N\geq 1$
supersymmetric
theories in any dimension.\cite{wit3}
This index has been very useful in providing criteria for
supersymmetry breaking.

For an $N=2$ theory in two dimensions, we have
a conserved $U(1)$ charge $F$ (the `fermion number' of
(\ref{algebra})), and in \cite[x]{cfiv}
Cecotti et al.
show that the `new index' $\,\,{\rm Tr}\, F(-1)^Fe^{-\beta H}\,\,$
depends only on $F$-term perturbations.\footnote{And thus is not an index
in
the
mathematical sense.  Rather it is related to what the mathematicians call
`holomorphic torsion.'}
This can also be thought of as a path
integral on the cylinder, now written in an $x^0$ as time operator
formalism.

The new index is actually a matrix since the boundary conditions at
spatial
infinity can be any vacuum of the theory. Let the left vacuum be $a$ and
the
right one $b$, and consider the matrix elements
\eq\label{newindex}
    Q_{ab} ={{i\beta}\over{L}}{\Tr}_{ab}\;(-1)^F\; F\;e^{-\beta H}.
\en
In \cite[x]{cfiv} it is shown that the matrix $Q$ is imaginary and
hermitian,
and that
\eq\label{metrel}
      Q_{ab} 
             = i(\beta g{\del_\beta}g^{-1} + n)_{ab}
\en
where $n$ is the coefficient of the chiral anomaly.
This expression comes from writing out the path integral calculation and
reinterpreting it in the $x^1$ as time operator language.
The fermion number becomes chiral charge,
and a relation between this and the stress tensor is used to get
$\del/\del\beta$.

This quantity is
particularly suited for extracting the soliton spectrum and other low
temperature properties of the model.
The simplest case is a model with a mass gap; clearly $Q$ will be
exponentially
small in $\beta m$ and typically each of the leading terms in an expansion
in
$\exp -\beta m$ is the contribution of a single massive
particle saturating the Bogomolnyi bound $m=|\Delta|$.%
\cite{fend,cfiv,witolive}

For our purposes, it has the additional advantage that
its definition does not
involve the precise normalization of the ground states
$\ket{i}$, which simplifies its computation.

\sect{The \sCP\ sigma models}

Non-linear sigma models define maps from spacetime into a riemannian
target manifold $M$.  Supersymmetric $d=2$ sigma models exist for any
target
manifold.  If the target manifold and metric is K\"ahler, the model will
be
$N=2$. \cite{zu}
(For a review, see \cite[x,y,w]{ag,novi,pere}).

\noindent A manifestly $N=2$ invariant superspace Lagrangian
is
\eq
{\cal L}=\half\int dx\;d^2\theta \;d^2\bar{\theta}\;\;
K(\Phi,\Phi^{\dagger}).
\en
where $K$ is the K\"ahler potential
and $\Phi_i$ are complex chiral superfields
\eq
\Phi_i=\Phi_i(x,\theta,\bar{\theta})=\varphi_i(x)+\sqrt{2}
\epsilon_{\alpha\beta}\theta^{\alpha}\Psi_i^{\beta}(x)+
\epsilon_{\alpha\beta}\theta^{\alpha}\theta^{\beta}F_i(x)~.
\en
Any term in $K$ which is globally defined on $M$ is a $D$ term, and
conversely
two choices of $K$ for which the K\"ahler forms $J=dz^i \wedge d\bar
z^{\bar
j}\del_i{\bar{\del_{\bar j}}} K$ are in different complex cohomology
classes
differ by $F$ terms.

For \CP, $\dim H^{1,1}(M,\RR) = 1$ and the K\"ahler class is specified by
a
single parameter.  We can take
\eq\label{kahler}
K(\Phi,\Phi^{\dagger})={1\over{g^2}}
\log(1+\sum_{i=1}^n\Phi_i^{\dagger}\Phi_i).
\en

The supersymmetric ground states of an $N=2$ sigma model are in one-to-one
correspondance with the complex cohomology classes of the target space,
and by spectral flow so are the chiral primaries.
Using semiclassical techniques to compute the chiral ring, one finds it to
be a deformation of the classical cohomology ring:
instantons can contribute to correlation functions of the chiral
primaries.
In simple cases the possible contributions are determined by the
chiral anomaly.
For \CP, the classical cohomology ring is the powers of the K\"ahler form
$x$ allowed on a $2n$-dimensonal manifold, up to $x^{n}$.
The instanton changes the relation $x^{n+1}=0$ to
\eqn\label{chiral}
x^{n+1}=e^{-2\pi/g^2}.
\enq
One can introduce as well a coupling $\theta$ to control a
topological term, which weighs a configuration of instanton number $w$ by
a factor $e^{i\theta w}$.  This combines with $1/g^2$ to make
a chiral coupling ~$t_1=2\pi/g^2+i\theta$ as in section 2.

For many purposes, a more useful definition of the \CP sigma model is
provided
by a gauged $N=2$ model, which constructs \CP as a quotient of $\CC^N$
(let
$N={n+1}$):
\eq\label{lgform}
   {\cal L} = \int d^4\theta \left [ \sum_{i=1}^N \bar{S}_i
    e^{-V} S_i + {N\over{g^2}} V\right ] .
\en
$S_i$ are $N$ chiral superfields which become
the homogeneous coordinates on \CP.
We have introduced a factor of $N$ with the coupling $1/g^2$ which will
make
the $N\rightarrow\infty$ limit well defined. \cite{coleman}
$V$ is a real vector superfield, whose components become the many
auxiliary
fields of the following component form of the Lagrangian:
\eqn\label{compform}
{\cal L}&=&{N\over g^2}\bigl\{\ (D_\mu n^*_i)^\dagger
  (D_\mu n_i)+\bar{\psi}^i(i\Dslash+\sigma+i\pi\gamma^5)\psi_i- \no \\
        &&+(\sigma^2+\pi^2)-\lambda(n^*_in_i-1)
	+ \bar\chi n^*_i \psi^i + \bar\psi^i n_i\chi \ \bigr\}.
\enq
The superfields $S$ have complex components $n_i$ and $\psi_i$.
(We rescaled them by $\sqrt{N}/g$.)
The constraint $n^*_in_i=1$ is imposed by the Lagrange multiplier
$\lambda$;
the phase of $n$ and $\psi$ is gauged by $A_\mu$ (which appears in
$D_\mu=\del_\mu+iA_\mu$).
The fields $\sigma$ and $\pi$ implement 4-fermi interactions and by the
equations of motion are equal to $\bar\psi^i\psi_i$ and
${i}\bar\psi^i\gamma^5\psi_i$ respectively.
The fermionic auxiliary fields $\chi$ constrain the $\psi_i$ to be
tangent to \CP.  They will play a secondary role in
our considerations.
The action has an additional chiral $U(1)$ symmetry,
$\delta\psi_i = \gamma^5\psi_i$, and
$\delta(\sigma+i\pi) = -2i(\sigma+i\pi)$, which
in the quantum theory will be anomalous.

Since the fields $S_i$ appear quadratically, it is possible to integrate
them out exactly, at least in terms of a one-loop determinant:
\eqn\label{effact}
{\cal L}&=&{-N}\Tr\log(-D_\mu D^\mu + \lambda)
+ {N}\Tr\log(i\Dslash+\sigma+i\pi\gamma^5) \no
\\
        &&+{N\over g^2}(\sigma^2+\pi^2-\lambda)
	+ {\rm fermionic}.
\enq
The determinant can be regulated straightforwardly (e.g. by
Pauli-Villars) and by supersymmetry the divergences will cancel in the
result.
As a functional of the auxiliary fields, it can be evaluated quite
explicitly
for constant fields \cite{dadda} and then as an expansion in either the
amplitude or frequency of the fluctuations around this.
In the large $N$ limit this allows us to solve the model: the $N$ in front
of
the action means that the remaining integration over auxiliary fields can
be
done by saddle point, and calculations at leading order in $1/N$ can be
done by
classical techniques.  For example, an S-matrix element would be given by
a sum
of tree diagrams; for a specified number of external particles these are
finite
in number.

Writing out the low energy effective action makes the physics of the model
clear.\cite{wit2,affleck,others}
At zero temperature we have an effective potential
\def\effpotzero{
\eqn
V_{eff}&=&{1\over {g^2}}(\sigma^2+\pi^2)
+{1\over {4\pi}}(\sigma^2+\pi^2)\log{\sigma^2+\pi^2\over\mu^2}
\no \\
&&-{1\over{g^2}}\lambda -{\lambda\over {4\pi}}\log{\lambda\over\mu^2}.
\enq
}
\effpotzero
It exhibits the dimensional transmutation in this asymptotically
free theory as only the combination $m_0\equiv\mu \exp -2\pi/g^2$ appears.
The effective potential has a minimum at non-zero
$\sigma^2+\pi^2\equiv m_0^2$
and an extremum at $\lambda=m_0^2$.
These give masses to the $n$ and $\psi$ particles, which are equal by
supersymmetry.
There will be kinetic terms induced for the auxiliary fields:
to lowest order, with the anomaly,
\def\loweff{
S_{eff}&=& N \int d^2x
{1\over 8\pi \mass^2} \left(F_{\mu\nu}^2 + (\partial_\mu \sigma)^2
+ (\partial_\mu \pi)^2\right)
+ {i\over 2\pi} \epsilon^{\mu\nu} F_{\mu\nu} \im\log (\sigma+i\pi)
+ V_{eff} + \ldots
}
\def\mass{m_0}
\eqn\label{eqloweff}
\loweff
\enq
Supersymmetry
tells us to expect a multiplet of particles associated with the $\sigma$,
$\pi$ and $\chi$ fields.
This is the point at which we see a much-noted analogy with QCD.
The vev for $\sigma+i\pi$ breaks chiral $U(1)$ spontaneously, suggesting
that
its phase is a Goldstone boson.  However this suggestion cannot be right
in
two dimensions (it also contradicts the expectations from supersymmetry)
and
for finite $N$ one is happy to find that just as in QCD, instanton effects
explicitly break this $U(1)$ to $\ZZ_N$.
This leaves a bit of a puzzle in that in terms of the coupling in
(\ref{compform}), rescaled as appropriate for the large $N$ limit, the
instanton action is $\exp -N/g^2$ and instantons should be invisible in
the
limit, so what eliminates the Goldstone boson?
The answer to this puzzle is that the role of the instanton in the story
was to
provide a field configuration in which the integrated anomaly,
$\int d^Dx F^{D/2}$ was non-zero despite being integral of a total
derivative.
In two dimensions we do not need the instantons -- a typical gauge
potential
grows linearly in space and $\int d^2x F$ will be non-zero without their
help.
Rather we must treat the anomaly on the same footing with the other
induced
kinetic terms, and the result is more analogous to the massless Schwinger
model:
the $\pi$ boson is massive (as are the other particles in the
supermultiplet),
and the gauge field is screened (so unlike the bosonic \CP model, the $n$
and
$\psi$ particles are not confined.)

There are still degenerate vacua labeled by the phase of $\sigma+i\pi$.
Although the phase is not quantized, the difference between the phases at
$x^1=\pm L$ is quantized.  This follows from Gauss' law and the screening
of
the gauge field.\cite{wit2}
We have
\eq
{1\over 4\pi m_0^2}\partial_1 F
-{i\over 2\pi m_0} \partial_1 \pi = {1\over N} J^0
\en
where $J^0$ is the electric current of the elementary fields $n$ and
$\psi$.
Integrating $\int dx^1$ and realizing that because of the screening,
$F$ vanishes at infinity, shows not only that the phase difference is
quantized
in units of $1/N$ but that we should think of the elementary particles as
solitons, in the sense that the associated field configuration
interpolates
between different choices of vacuum.

The conclusion is that we
should think of the chiral $U(1)$ as being explicitly broken to $\IZ_N$
just as for finite $N$, and the further spontaneous breaking of $\IZ_N$ is
associated with multiple vacua and solitons.  This allows us to identify
the concepts of section 2 in this language.
The ground states are characterized by the phase of $\sigma+i\pi$, so we
can
associate the choice of vacua $a$ and $b$ in (\ref{newindex}) with
$\sigma+i\pi\rightarrow_{x^1\rightarrow-L} m_0\exp 2\pi i  a/N$
(resp. $b$ and $+L$).
Clearly it is a function only of $a-b$.  Furthermore,
this difference must be $O(N^0)$,
because we have a lower bound $m_0|a-b|$ on the energy in this sector.
Thus a representative choice is $\pi\rightarrow \pm 2\pi m_0 a/N$,
$\sigma=Om_0+O(1/N^2)$ and we can think of a ground state as labelled by a
value of $\pi$.
This is only one possible basis and it is not in fact the basis defined by
spectral flow; we will see this shortly.

One can also write a manifestly $N=2$ supersymmetric
effective Lagrangian.\cite{dadda}
A priori this is a functional of $V$, but by gauge invariance the
effective
action should depend
only on the field-strength superfields $X$ and $\bar{X}$
($V$ is not gauge invariant):
\eq
  X=D_L\bar{D}_R V,\;\;\;\;\;\bar{X}=D_R\bar{D}_L V
\en
\eq\label{Xfield}
X=(\sigma+i\pi) + \bar\theta_L\chi_R + \chi_L\theta_R
+ \bar\theta_L\theta_R(\lambda-F)|_{x_{ch}}
\en
The effective action then is
\eqn
S_{\rm eff}={N\over{2\pi}}\!\int\!\!d^2x\left\{
\int \!\! d^2\theta W(X)\!+\!\!
\int\!\! d^2\theta \bar{W}(\bar{X})
\!+\!\!\int\!\! d^4\theta[Z(X,\bar{X},\Delta,\bar{\Delta})]\right\}
\enq
where
\eq
W(X)={1\over{2\pi}} X({1\over N}\log X^N-1+A(\mu)-i\theta)
\en
\noindent where $A$ is a renormalized coupling.
(For more details, see \cite[x,y]{dadda,cv})

Now many $N=2$ supersymmetric theories in two dimensions admit
a Landau-Ginsburg description, meaning they can be described by a
superspace Lagrangian of the form
\eq\label{lglang}
{\cal L}=\int d^4\theta \sum_i \phi_i\bar{\phi}_i +\int d^2\theta
W(\phi_i)+h.c.
\en
The superpotential $W$ is an analytic function of the complex superfields.
Since there is a non-renormalization theorem
for it, one can directly infer that
the ground states of the theory are $dW(\phi)=0$.
The chiral ring is the ring of polynomials generated by the
$\phi_i$
modulo the relations
$dW(\phi)/d\phi_i = D \bar D \phi_i \sim 0$.
(For a review, see \cite[x]{lvw}.)

Since our \CP\ effective action now has the form of a Landau-Ginsburg
theory,
it follows\cite{cv} that its chiral ring is the powers of $X$ mod
\eq
X^N = \exp -A+i\theta \equiv m_0^N.
\en
This is an alternate derivation of the chiral ring (\ref{chiral}).
It would be interesting to get the $N\rightarrow\infty$ results below in a
manifestly supersymmetric way.

The ground states of our previous section are the solutions of $W'(X)=0$,
in other words $X=m_0 \exp 2\pi i n/N$.  Clearly these are eigenstates
under
multiplication by the chiral ring, and therefore not the states defined by
$\ket{j}=X^j\ket{0}$.
By considering the $\IZ_N$ symmetry, one sees that these are
the conjugate states
\eq\label{constates}
\ket{j} = m_0^j N_j \sum_k e^{2\pi i (j-j_0) k/N}
\ket{\sigma+i\pi=m_0 \exp 2\pi i k/N}
\en
up to a phase $j_0$ and normalizations $N_j$
not determined by this argument.
To get these we should make the spectral flow
argument explicit, and we will discuss this below.

\sect{The ground-state metric for \sCP}

In reference \cite[x]{cvsig}, the $tt^*$ equations are given for
the supersymmetric  \CP\ model on a K\"{a}hler manifold $M$.
As we have just seen, the chiral ring is generated by a single
element $x$ with the relation
\eq\label{cpnring}
   x^N=t^N
\en
where $t=m_0 e^{i\theta}$ is a complex chiral coupling as in section 2.
\comments{The topological metric $\eta_{ij}=\langle
x^i|x^j\rangle=\delta_{i+j,n-1}$ in the basis corresponding to
$\int x^{n-1}=1.$}

The action can be written in the following form
\eq
S=-{N\over 4\pi}\ln t\int d^2yd^2\theta \; x + {\rm c.c.}
\en
where  $x=x(\Phi_i,\bar{\Phi}_i)=
\bar{D}D\ln(1+\sum\Phi_i\bar{\Phi}_i)$ represents
the K\"ahler class, $\ln t~ x$ is the K\"ahler form and
$\Phi_i,\bar{\Phi}_i$ are chiral superfields.

\noindent To write down the $tt^*$ equations,
we need to find the operator corresponding to a perturbation of $t$,
and its action on the chiral ring $(X^n,\ldots,X,1)$.
It is represented by the matrix
\[ C_t
={N\over4\pi t}\left(\eqa{cccccc}
0 & 1 & 0 & \dots & 0 & 0 \\
0 & 0 & 1 & \dots & 0 & 0 \\
\dots & \dots & \dots & \dots & \dots & \dots \\
\dots & \dots & \dots & \dots & \dots & \dots \\
0 & 0 & 0 & \dots & 0 & 1 \\
t^N & 0 & 0 & \dots & 0 & 0
\ena \right) \]

\noindent The $\IZ_N$ symmetry implies that the metric
$g_{i\bar{j}}=\langle\bar j|i\rangle$ is diagonal.
Thus, defining
\eq
         q_i  =  \ln g_{i\bar{i}} \qquad\qquad
          g_{i\bar{j}}  =  \langle\bar j | i\rangle\qquad\qquad
          q_{i+N} \equiv 2N\log|t| + q_i
\en
the $tt^*$ equations  are
\eq
    {16\pi^2 t^2\over N^2\beta^2} \partial_t\partial_{t^*}q_i +
          e^{(q_{i+1}-q_i)} -
        e^{(q_i-q_{i-1})}=0.
\en
The metric $g$ is a function only of $|t|^2$, because it is a path
integral
with total
chiral charge zero, and chiral charge non-conservation is proportional to
instanton number.
Thus the equations become o.d.e.'s in terms of $|t|$.
We can write them in terms of the dimensionless parameter
$x=\beta t/2\pi$, but to do this we need to take out the dimensional
factors
in $g_{i\bar j}$ coming from the definition (\ref{constates}).
Thus we redefine
\eq
         q_j  =  \ln g_{j\bar{j}} + 2j\log|t| \qquad\qquad
          q_{j+N} \equiv q_j.
\en
With this straight, we will use $x$ as our coupling, and call it $\beta$
in the following.  The $tt^*$ equation  becomes
\eq\label{ttstar}
    {4\over N^2} \partial_\beta\partial_{\beta^*}q_i +
          e^{(q_{i+1}-q_i)} -
        e^{(q_i-q_{i-1})}=0.
\en
This equation is the affine $\hat{A}_{n}$ Toda equation.
(In \cite[x]{cvsig} it is shown that on general grounds
$q_i+q_{N-i-1}=0$, which reduces
the equation to the $C_m(BC_m)$ Toda equation with $n=2m(n=2m+1)$,
but we will not use this.)

A solution should be determined by
the boundary conditions near $\beta\sim 0$ and $\beta\sim \infty$.
In \cite[x]{cvtop} these are found explicitly for the small $\beta$ limit
by a
semiclassical calculation of the metric.  For the large $\beta$ limit it
would
suffice to know (on general grounds) that the solution is exponentially
small
in $\beta$; in fact the precise form of the leading exponential is
determined
by the soliton spectrum, which is already known for these (integrable)
models.

For the cases of $\CC P^1$ and $\CC P^2$, the $tt^*$ equations become
special cases of the Painlev\'e ${\rm I\!I\!I}$ equation, for which the
connection formula between small and large $\beta$ asymptotics is known.
A more general discussion is given in \cite[x]{cv}.

A reasonable ansatz for the large $N$ limit would be that the metric and
index
are continuous functions of the variable $s\equiv i/N$.
We will verify that this is true for the boundary conditions of
\cite[x]{cvsig};
it will also follow from the explicit calculation in section 5.
Computing the metric is very similar to computing the free energy with
specified boundary conditions at $x^1\rightarrow \pm L$,
which would produce $\exp N S_{eff}$ at an
appropriate saddle point.  With this motivation we redefine
\eq
      q_j  = {1\over N} \log g_{j\bar{j}} + 2{j\over N}\log|t|
\en
The $tt^*$ equation becomes:
\eq
   {4\over N} \partial_{\beta}\partial_{{\beta}^*}q_i +
        e^{N(q_{i+1}-q_i)} - e^{N(q_i-q_{i-1})}=0
\en
with $q_{i+N}=q_i$.
We see that the $N$ dependence is consistent with $q(\beta,s)$ having a
good
large $N$ limit, satisfying ($q'={{\partial q}\over{\partial s}}$)
\eq
 4\partial_{\beta}\partial_{{\beta}^*}q + {{\partial}\over{\partial s}}
e^{q'} =
0.
\en
Defining $H=q'$,
\eq\label{oureqn}
       4\partial_{\beta}\partial_{\beta^*}H +{{\partial^2}\over
      {\partial s^2}} e^H = 0.
\en

This equation has been studied in several contexts.
It was first noted for a connection with 4D self-dual
gravity.\cite{bf,gd,eh}
More recently, it has been
studied in the context of the large $n$ limit of $W_n$
algebra.\cite{bakas,bg,park}
It is also a well known scaling limit of the two-dimensional
infinite Toda lattice.\cite{saveliev}
A formal solution of the boundary value
(Gours\'at) problem for the equation has been given
in \cite[x,y]{kssv}.

We still need to specify the boundary conditions to select a solution to
the equation.  One can take the large $N$ limit of Cecotti and Vafa's
boundary
conditions at large and small $\beta$; they will follow independently
from the results of section 5 so we will just quote them here.
The $s$ boundary conditions are $H(\beta,s)=H(\beta,s+1)$.

We can deduce the large $N$ limit of our metric for small $|\beta|$
from the semi-classical result of Cecotti and Vafa.\cite{cvsig}
This is essentially the two-point function $\vev{\phi_i\bar\phi_j}$
reduced to constant field configurations, or
$\int d\varphi~x^i \wedge *x^j$.
They find
\eq
 g_{r\bar{r}} ={{r!}\over{(n-1-r)!}} [|\beta|(-\ln (|\beta|/2)-\gamma)]
^{n-1-2r}
\en
where $\gamma$ is Euler's constant, a factor predicted by the connection
formula for the $n=1,2$ equations, and described in \cite[x]{cvsig} as
a one-loop correction to the semiclassical calculation.
\noindent This becomes, in the large $N$ limit
\eq
\label{smallbeta}
e^H=
{s(1-s)\over|\beta|^2(-\ln (|\beta|/2) - \gamma)^{2}}
\en
(for $0<s<1$ and defined elsewhere by periodicity).

Actually, this is already an exact solution to (\ref{oureqn})
(this does not depend on the value of $\gamma$).
At finite $N$ perturbative calculations around the trivial
background are one-loop exact; the small $\beta$ boundary condition failed
to be a solution because of instanton corrections.
In the large $N$ limit these in some sense become trivial.
We saw in section 3 that in terms of the rescaled coupling the
instanton weight is $\exp -2\pi N/g^2$ and that we can
understand a lot of physics even if we call this zero.
Here the instantons are responsible for the boundary condition
$H(\beta,s)=H(\beta,s+1)$.
Whether this prevents (\ref{smallbeta}) from being a solution depends
sensitively on how we treat the region $s=0$, since (\ref{smallbeta})
has a kink there.
One prescription which makes sense is to solve not imposing
$H(\beta,s)=H(\beta,s+1)$ but allowing arbitrary $s$
dependence, and if the answer satisfies $H(\beta,0)=H(\beta,1)$, accept
it.
If we use this definition
we cannot attribute the corrections to the small $\beta$ limit
to instantons.
An analogous problem was studied in \cite[x]{affleck}, that
of understanding the theta dependence of the large $N$ bosonic \CP\ model.
There instantons were also unimportant and the non-perturbatively
small soliton action $S\sim\beta \mu e^{-1/g^2}$
controlled the theta dependence.

For large $\beta$, each sector with one soliton of mass $m $ satisfying
the
Bogomolnyi  bound contributes $((f+1)-f)\exp -\beta m$ times a factor
depending
on its central charge $\Delta$ to the new index,
and by (\ref{metrel}) to $H$.
{}From section 3 we see we have $N$ such solitons with mass $m=m_0$; the
central charge is $\pi(+L)-\pi(-L)$ or one can just linearize
(\ref{oureqn}) to
see the appropriate boundary condition
\eq
\label{largebeta}
 H(s)\sim -{{\exp{(-2\pi|\beta|)}}\over{\sqrt{2\pi|\beta|}}}
\cos (2\pi s)
\en
which satisfies our equation to  first order.
This limit is not a solution and one could use (\ref{oureqn}) to generate
corrections to $H$ coming from multi-soliton sectors.

One might at first say that in the $tt^*$ formalism, the existence of the
solitons is fed in through the large $\beta$ boundary condition.
However even without this
it was clear that some physics must modify the solution
(\ref{smallbeta}) -- it is singular at $|\beta|=2\exp -\gamma$.
(Classically, without $\gamma$, this would be the `zero volume limit of
the
target space.')
In \cite[x]{cvtop,cvsig,cfiv} it was typically found that requiring
regularity on solutions of the $tt^*$ equations was a strong constraint,
which to some extent predicted physical boundary conditions.
In this sense quantum field theoretic information seems to be emerging
from the formalism in a rather mysterious way.
We do not know enough about general solutions of (\ref{oureqn}) to make
a strong statement here, but we are certainly seeing some form
of this novel way to predict non-perturbative corrections.

If we write
\eq
e^{H} = -R\ s(1-s)\ e^{\varphi(\beta,\bar\beta)}
\en
(\ref{oureqn}) reduces to the Liouville equation for a 2d metric with
constant curvature $R$.
Its solutions
are related by Legendre transform (as we will see in section 5)
to self-dual Einstein metrics,
and with this motivation, this ansatz was considered in \cite[x]{gd}.
The $R>0$ case is related to the
Eguchi-Hanson metric, a gravitational instanton.\cite{eh}
An `elliptic' $R<0$ solution can be related to a similar
(but singular) metric.\cite{gd}
Our $R<0$ solution (\ref{smallbeta}) is the `parabolic'
case.\cite{seiberg}

\sect{Finite temperature results and the new index}

Recall the effective action (\ref{effact}).
It has been
extensively studied at $T=0$ \cite{dadda} and at finite temperature
(typically not in the supersymmetric context, but the results can be
easily
adapted).
\cite{davis,affleck}
In the large $N$ limit it is $O(N)$ and we can calculate bulk quantities
like
the free energy simply by extremizing it with respect to the auxiliary
fields.
The `new index' is computed similarly, with the main differences being
that
we take periodic fermi boundary conditions (and have unbroken
supersymmetry),
we insert the fermion number operator $F$ (this will be done by
differentiating
with respect to a coupling at the end), and we fix the boundary conditions
at
$x^1=\pm L$ to go to two possibly different values of
$\sigma+i\pi$.
This last condition means that we need to consider non-constant background
fields in the functional integral.
For general background fields this is quite complicated, but what saves us
is
that the required variation is small, of $O(1/N)$, so we only need
the
leading terms in an expansion in derivatives and amplitude.
The derivative terms (to the accuracy we need them) are
\def\mass{\sigma}
\eqn\loweff\enq
where we no longer assume $\vev{\sigma}=m_0$.
The finite temperature effective potential is
\eqn
V_{eff}&=&{1\over {g^2}}(\sigma^2+\pi^2)-
\sum_{k^0} \int{{dk^1}\over{(2\pi)^2}}{\rm
tr}
\ln [k_\mu\gamma^\mu-A_\mu\gamma^\mu-(\sigma
+i\pi\gamma^5)]\no \\
&&-{1\over{g^2}}\lambda +\sum_{k^0} \int {{dk^1}\over{(2\pi)^2}}
\ln [(k_\mu-A_\mu)^2+\lambda] \label{kzerosum}\\
&=& V_{eff}|_{T=0} \no\\
&&+\int^{\infty}_0{{dk}\over{2\pi}} \ln|1-e^{-\beta\sqrt{k^2+\sigma^2}}
e^{i\beta A_0}|^2\label{fermterm}\\
&&-\int^{\infty}_0{{dk}\over{2\pi}} \ln|1-e^{-\beta\sqrt{k^2+\lambda}}
e^{i\beta A_0}|^2.
\enq
This is essentially the standard expression from statistical mechanics of
a
free field \cite{kapusta} (with chemical potential $iA_0$)
with one difference: we incorporated the periodic fermion boundary
conditions,
which led to the sign change in (\ref{fermterm}).

The new index is
\eq\label{index}
    Q_{ab} ={{i\beta}\over{L}}{\Tr}_{ab}\;(-1)^F\; F\;e^{-\beta H}.
\en
where $a$ and $b$ characterize the vacua at spatial infinity.
We can rewrite it as a path integral with an insertion of
the fermion number charge $\int dx^1 J^0_F$.
If we introduce a new gauge
field $B_\mu$ which replaces $A_\mu$ in coupling to the bosons,
the fermion number will be given by
differentiating $dV_{eff}/dA_0$ before imposing $B_\mu=A_\mu$.
Thus we will split $S_{eff}$ into two parts, $S_B(\lambda,B)$
from the integral over the $n_i$, and $S_F(\sigma,\pi,A)$ from the
$\psi_i$.
All the derivative terms of (\ref{eqloweff}) are in $S_F$.
We know that $H=0$ on our ground states, so $Q$ will just be
the expectation value of $F$.

It may sound a bit strange to be extremizing a Euclidean action which one
might
have thought should be non-negative.  The reason it need not be (and is
not)
bounded below is that it depends on a Lagrange multiplier, $\lambda$.
There can be several extrema, so we first minimize the fermion effective
action, then determine $\lambda$ by supersymmetry.  $S_B$ enters only in
computing $S_{eff}=0$, and we will not write this out in the following.

The boundary conditions $ab$ select supersymmetric vacua.
As we saw in section 3, these are determined by expectation values
$\im\log(\sigma+i\pi)=2\pi a/N$ and $2\pi b/N$, so $Q$ will be a function
of
$p=(a-b)/N$.
We must also specify the other fields: they will be independent of $x^0$
and satisfy the equations of motion: in $A_1=0$ gauge,
\eqn
-\partial_1^2 A_0 - i \partial_1\pi &=& {\delta V_{eff}\over\delta A_0}\\
-\partial_1^2 \pi + i \partial_1 A_0 &=& 0.\label{eqpi}
\enq
The general solutions are exponentials, but there are special solutions.
One which works independently of $V_{eff}$ is
\eq
\partial_1 A_0 = \partial_1^2 \pi =0.
\en
One can check that this preserves a supersymmetry
(up to $O(1/N)$ corrections).

We will substitute this solution directly into the effective action.
Thus we take $\pi=p \sigma x^1/L$ and $A_0$ constant, giving
\eqn\label{zeromodes}
S_F&=& N \int d^2x
{1\over 8\pi } (p/L)^2 - {i\over 2\pi} (p/L) A_0
+ V_F(\sigma,A_0) + \ldots\\
&=& \beta N L \left(
{1\over 8\pi } (p/L)^2 - {i\over 2\pi} (p/L) A_0
+ V_F(\sigma,A_0) + \ldots \right)
\enq
The anomaly $\int~\pi F$ contributes because we integrate
by parts and drop a boundary term (more on this below).
The expansion in derivatives of $\pi$ becomes exact in the limit
$L\rightarrow \infty$.
Using $dS_F/dA_0=0$ at the saddle point,
$Q = i\beta/L~ dV_F/dA_0 = -(\beta/2\pi)(p/L)$
at an extremum of $S_F$ in $\sigma$
and $A_0$, given $p$ and $\beta$.

The basis $\ket{a}$ is not the basis of (\ref{constates}).
There the basis was defined by acting on
a vacuum $\ket{0}$ with chiral fields of definite charge, $X^a$.
This is the conjugate basis and we have
\eq
Q(j,\beta) = \int dp~ e^{i (j-j_0) p} Q(p,\beta).
\en
Now we need $j_0$, which is determined by the spectral flow construction.
We have not done this construction in detail
but we believe the essential points are as follows.
We need to work on a disk, say with radial coordinate $x^1$ and
angular coordinate $x^0$, and boundary $x^1=1$.
The quantum number $j$ is the variable conjugate to the phase
of $\sigma+i\pi$, or working near $\pi=0$, conjugate to $\pi$.
This is not $\partial_1\pi$, because the anomaly $\pi\partial_1 A_0$
changes the symplectic structure.
Instead it is $j=N(\partial_1\pi-2 i A_0)/2$.
The ground state is clearly $j=0$,
but this is in the `Neveu-Schwarz' sector
(antiperiodic fermion boundary conditions on the cylinder)
and we must turn on a gauge field $A_0=1/2$ to
turn it into the corresponding supersymmetric ground state.
This shifts $j\rightarrow j-N/2$ and the relation to section 4
is $j-j_0=N(s-1/2)$.
We are then instructed to use this state (with norm $1$)
as a boundary condition on
our cylinder, which justifies dropping the boundary term in
(\ref{zeromodes}).

Since $j-j_0=N(s-1/2)$,
at fixed $s$ we can also do this integral by saddle point,
producing
\eq\label{indexr}
Q(s,\beta) = -{N\over 2\pi i}{d\over ds} {\beta\over
NL}S_F({\beta,s})|_{min}
\en
with
\eqn
	{\beta\over NL} S_F &=& -{1\over 2\pi}(2\pi (s-\half)+ \beta
A_0)^2
 + \beta^2\sigma^2 {1\over{4\pi}}\left(\ln{\sigma^2\over m_0^2}-1\right) -
\\
    &&~\beta\int^{\infty}_0{{dk}\over{2\pi}}
{}~~\ln \left|1-e^{-\beta\sqrt{k^2+\sigma^2}+i\beta A_0}
\right|^2.\no
\enq

Let $u \equiv y^2 \equiv \beta^2\sigma^2$, $A\equiv\beta A_0$,
$s'=s-1/2$  and
$\lambda=\log(\beta m_0/2\pi)$, so
\eqn\label{sfermi}
{2\pi\beta\over NL} S_F = -\lambda u - (2\pi s'+ A)^2
+\half u (\ln {u\over4\pi^2}-1) \\
-\int^{\infty}_0{{dk}}~\ln \left|1-e^{-\sqrt{k^2+u}+i A}
\right|^2\no
\enq

We will now show that the metric related to the index $Q$ by
(\ref{metrel})
indeed solves the `heavenly' equation (\ref{oureqn}).
We still need to minimize the effective
action with respect to $\sigma$ and $A_0$.
Although this cannot be done in closed form, nevertheless the minimization
procedure is natural in this context:
it amounts to a double Legendre transform of the effective action from
$(\sigma,A_0)$ to $(\beta,s)$.

We will start from a rather little known fact:
considered as a function of the background fields $\sigma$ and
$A_0$, the effective action $S_F$ satisfies a linear p.d.e.
(essentially the Laplace equation):
\eqn\label{xlaplace}
\left[y{\partial\over\partial y}{1\over y}{\partial\over\partial y} +
{\partial^2\over\partial A^2} \right] S_F(y,A) = 0.
\enq
Clearly the zero temperature part works.
At finite $\beta$ one can verify it by expanding the $\ln$ in
(\ref{sfermi})
and integrating termwise to get a sum over Bessel functions as in
\cite[x]{hw},\footnote{
This is a straightforward calculation.
To see it explicitly,
turn on the {\tt `expandedversion'} switch in the tex file,
hep-th/9312095.}
but it is much clearer
in terms of the sum over timelike momenta
(\ref{kzerosum}):
\eqn\label{kzerosumtwo}
F_0 = \sum_{k_0} \int {{dk_1}\over{(2\pi)^2}}
\ln [(k_0-A)^2+k_1^2+y^2].
\enq
Now (\ref{xlaplace}) will be true if
\eq
 v \equiv {1\over y}{\partial F_0\over\partial y} =
 \sum_{k^0} \int {{dk_1}\over{(2\pi)^2}}
{1\over (k_0-A)^2+k_1^2+y^2}
\en
satisfies the three dimensional Laplace equation
\eqn
\Delta v =
{1\over y}{\partial\over\partial y}y{\partial v\over\partial y} +
{\partial^2 v\over\partial A^2}   = 0.
\enq
Since this is linear we can verify it for each term in the sum,
and shift $A$ in each to absorb $k_0$.
\\
Now, as a function of $x^\mu\equiv(k_1,A_0,y \cos \theta,y \sin\theta)$,
the integrand is $1/x^2$ which solves the four dimensional Laplace
equation.
Thus the operator $\Delta$ on the integrand is equal to
$-\partial^2/\partial k_1^2$, which integrates to zero.

\expandedversion{
One can also see this in the low temperature expansion.
Consider the finite temperature part of $S_{eff}$,
call it $f_0$.
Then,
\eq\label{intfzero}
 {1\over y}{\partial f_0\over\partial y} =
\int^{\infty}_0{{dk}\over{\sqrt{k^2+y^2}(e^{\sqrt{k^2+y^2}- i
A}-1)}}\,+\,\int(A\rightarrow -A)
\en
Following \cite[x]{hw}, we expand
\eq
{1\over{e^{\sqrt{k^2+y^2}- i
A}-1}}=\sum^{\infty}_{p=1} e^{ipA-p\sqrt{k^2+y^2}}
\en
and integrate termwise
\eq
\int^{\infty}_0 dx {e^{-p\sqrt{x^2+y^2}}\over{\sqrt{x^2+y^2}}}=K_0(py)
\en
to find
\eq
 {1\over y}{\partial f_0\over\partial y} =\sum^{\infty}_{p=1}
\cos{A p}~~K_0(py)
\en
This is clearly a solution of the Laplace equation
\eqn
y\left[{1\over y}{\partial\over\partial y}y{\partial\over\partial y} +
{\partial^2\over\partial A^2} \right]
{1\over y}{\partial f_0\over\partial y}= 0.
\enq
Integrating once produces (\ref{xlaplace}).}
\else
The sum over $k_0$ in (\ref{kzerosumtwo}) of course does not converge, so
to
use this we would need to subtract the zero temperature part.
\fi
The result is that the finite temperature part $f_0$
(the integral) in (\ref{sfermi}) satisfies (\ref{xlaplace}).
Let us combine it with some of the zero mode terms in (\ref{sfermi}),
defining
\eq\label{ff}
f(u,A) = -A^2 +\half u \ln u - f_0(u,A).
\en
Now we do a Legendre transform from $A$ to $s$ which minimizes
the action in $A$
\eqn
K(s',u) = f(u,A) - 4\pi s' A
\enq
and we have
\eq\label{der1}
s' = {1\over 4\pi}{\partial f\over\partial A}~~~~~~~~
A = -{1\over 4\pi}{\partial K\over\partial s'}~~~~~~~~
{\partial K\over\partial u}={\partial f\over\partial u}.
\en
Using this we turn  (\ref{xlaplace}) (applied to $f$) into
\eqn\label{pleb}
0 &=& 4u{\del\over\del u}[{{\del f}\over{\del u}}] +
{{\del^2 f}\over{\del A^2}}\no\\
 &=& 4u\,d\;({{\del f}\over{\del u}})\wedge
dA-d\;({{\del f}\over{\del A}})\wedge du\no\\
&=& u\,d\;({{\del K}\over{\del u}})\wedge d\;({{\del K}
\over{\del s'}}) + 4\pi^2 ds'\wedge du
\enq
This last equation is essentially Pleba\'nski's equation\cite{pleban}
\eqn
{{\del^2 K}\over{\del u^2}}{{\del^2 K}\over{\del s'^2}}-
\left({{\del^2 K}\over{\del u\del s'}}\right)^2 ={4\pi^2\over u} .
\enq
We now perform a second Legendre transform from $u$ to $\lambda$
which will minimize in $u$
\eq
4\pi^2 J(s',\lambda)= K(s',u)-(2\pi s')^2 -\lambda u
\en
with
\eq\label{der}
\lambda = {{\del K}\over{\del u}}~~~~~~~~~
u = -  4\pi^2{{\del J}\over{\del \lambda}}
\en
This turns (\ref{pleb}) into
\eqn
(-{{\del J}\over{\del \lambda}})\, d\lambda\wedge d\;({{\del
J}\over{\del s'}}+2s') &=& -  ds\wedge d\;(-{{\del J}\over{\del
\lambda}})\\
{\partial^2 J\over\partial s'^2} + 2&=& {\partial \over\partial
\lambda}\log\left(-{\partial J\over\partial \lambda}\right)\label{almost}
\enq
Remembering
the result for the index (\ref{indexr}), and using the relation
(\ref{metrel}),
\eqn
{\partial^2 J\over\partial s'^2}
&=&  -{i\over  N}{\partial Q\over\partial s'}\\
&=& {\partial H\over\partial \lambda}
\enq
Using this in the l.h.s. of (\ref{almost}) and integrating once $d\lambda$
gives
\eqn
 H+2\lambda = \log\left(-{\partial J\over\partial
\lambda}\right).
\enq
There is a constant of integration which is determined
by consistency, for example by considering
the large $\beta$ limit of (\ref{uresult}).
It eliminates a term $-u(1/2+\log 2\pi)$ in (\ref{sfermi}).

\noindent Finally differentiate $d/d\lambda$
(\ref{almost}) and substitute to get
\eqn
e^{-2\lambda}{\partial^2 \over\partial \lambda^2} H +
{\partial^2 \over\partial s^2} e^H = 0.
\enq
\noindent Remembering $\lambda=\log x$,
this is our equation (\ref{oureqn})
for the $tt^*$ metric.
We also find
\eq\label{uresult}
u= 4\pi^2 e^{H+2\lambda}.
\en

Readers familiar with the work of Boyer and Finley\cite{bf} and
Gegenberg and Das\cite{gd} will recognize
that we are applying methods developed in the study of solutions of
the complex vacuum Einstein equations with a self-dual metric
admitting at least one Killing vector field.
They relate (by Legendre transform) the `heavenly'
equation (\ref{oureqn}) to Pleba\'nski's equation
\cite[x,y]{pleban,bf} reduced by symmetry with respect to
a so-called `rotational' or `non-KSD' Killing vector field, i.e.
a Killing vector whose covariant derivative has a non-zero
anti-self-dual part.
Our solution does not depend on the phase of $\beta$ (the theta
parameter),
which implies that the self-dual metric will have two rotational Killing
vectors.
Both papers suggest that such a self-dual metric must have a
`translational'
or `KSD' Killing vector, i.e. one whose covariant
derivative is self-dual.  All such self-dual metrics can be obtained by
Legendre transform of a solution to the Laplace equation, 
thus we could expect that our solution to the
`heavenly' equation could be obtained
by performing two Legendre transforms on a solution to the
Laplace equation.
This expectation was helpful to us, though we note that the determination
of
the index as the double Legendre transform of the effective potential is
really a consequence of the physical definition of the index starting
from (\ref{lgform}) and not these more abstract considerations.

\medskip
Is the correct solution completely determined from
$tt^*$ considerations?  The elliptic nature of (\ref{oureqn}) means
that if the boundary conditions are correct and `reasonable' then we might
expect its solution to be unique.
Thus we want to compare the limits $\beta\rightarrow 0$ and $\infty$ of
this
solution with the boundary conditions determined by semiclassical
considerations in \cite[x]{cvsig} and quoted in section 4.
The large $\beta$ limit is easy, by expanding (\ref{sfermi}) in
$e^{-\beta\sigma}$ with $\beta\sigma=y=2\pi\beta+O(e^{-\beta\sigma})$.
For small $\beta$ we want to make contact with (\ref{smallbeta}),
\eqn
e^H=
{1-4s'^2\over4|\beta|^2(-\ln |\beta| - \gamma + \ln2)^{2}}.\no
\enq
The inverse of the above Legendre transforms
can be performed analytically in this limit
and are essentially the ones done in \cite[x]{bf,gd}.$^{\S}$

\expandedversion{
Let $\lambda'=\lambda+\gamma - \ln2$ and $a=1/2$, then we have
\eq
\log(-{{\del J}\over{\del\lambda}})=\log\left[{{a^2-s'^2}\over{
\lambda'^2}}\right]
\en
or
\eqn
{{\del}\over{\del\lambda}}\log(-{{\del J}\over{\del\lambda}})
&=&-{2\over{\lambda'}}\\
&=& 2 + {{\del^2 J}\over{\del s'^2}}
\enq
This implies
\eq
J(\lambda,s')= \left[{a^2-s'^2\over{\lambda'}}\right]
 - s'^2
\en
Using (\ref{der})
\eq
u=4\pi^2{{a^2-s'^2}\over{\lambda'^2}}
\en
which by inversion gives
\eq
\lambda' = 2\pi{{\sqrt{a^2-s'^2}\over{\sqrt{u}}}}  =  {{\del
K}\over{\del u}}+\gamma - \ln2
\en
which determines $K(s,u)$ to be
\eq
K(s,u)=4\pi\sqrt{u(a^2-s'^2)}+u(\log2 -{{\gamma }})
\en
Using (\ref{der1})
\eq
{{\del K}\over{\del s}}=-4\pi{{s'\sqrt{u}}\over{\sqrt{a^2-s'^2}}} =
-4\pi A
\en
This can be inverted to find $s'$ in terms of $A$,
and replacing $s'$, we find $f$ from (\ref{ff}).
}\fi
\noindent The results are that in this limit
\eqn\label{smleg}
u=\pi^2{{1-4s'^2}\over{(\lambda+\gamma - \ln2)^2}}
\qquad\qquad
s'={1\over 2}{A\over{\sqrt{u+A^2}}}.
\enq
and the function defined in (\ref{ff}) has limit
\eqn
f = (\log2- \gamma) u - 2\pi\sqrt{u+A^2}.
\enq

This is to be compared with the high temperature limit of the effective
action (\ref{kzerosum}).  This is a one-dimensional limit and the
leading term is obtained
by keeping only the dominant term in the sum over frequencies
$k_0$ (and losing explicit periodicity in $A_0$).
The subleading terms are harder but fortunately are known:\cite{hw}
\eq\label{d3}
f_0= 2\pi\sqrt{u+A^2} + \half u\log u - A^2 +
(\gamma -\log4\pi-\half)u + \ldots
\en
and after combining with the zero temperature terms,
we see that the boundary conditions agree.

There is a technical point which would have to be addressed
to actually prove that the
solution is uniquely determined by the (asymptotic)
semiclassical boundary conditions.
This is the effect of the kink at $s=0$ in (\ref{smallbeta}),
which is not present in the true solution.
One would hope that the equation
(\ref{oureqn}) is stable under such a perturbation.

The minimization procedure brings with it the possibility of a phase
transition.  Since the \CP\ model has a
$\IZ_N$ symmetry restoration transition
at finite temperature,\cite{davis}
this would seem quite possible.  On the other hand the
bosonic sector has no transition and thus with periodic boundary
conditions the fermionic effective action will not either.  However,
in a sector with finite soliton number, $A_0\ne 0$ so it is not a priori
obvious that a transition is impossible.
If the $\IZ_N$ symmetry restoration transition
were present at some $\beta=\beta_c$,
we would see $u=\beta^2\vev{\sigma}^2=0$ there,
and for $\beta<\beta_c$
we would expect to see multiple extrema with $u=0$ and $u<0$.
This is not present in the $\beta\rightarrow 0$ limit (\ref{smleg})
and we conclude that this transition is not present in the new index.
Similarly, TBA calculations of the new index
for finite $N$ could have shown phase transitions,
but did not.\cite{intril}

There should be a similar direct computation of $g_{i\bar j}$ itself.
This would require doing the spectral flow construction to find
the correct normalization of the ground states (\ref{constates}).
The main point here may be the following:
given $\ket{0}$, we do not need spectral flow
but can apply chiral fields to the Ramond vacua
to produce $\ket{i+1}=X\ket{i}$.
The lowest component of $X$ is $\sigma+i\pi$ and since $\vev{\sigma}$
depends on the couplings, so do the normalizations.
The point is that $\vev{\sigma}$ depends on the state $\ket{i}$.
If we build up $\ket{i}$ one step at a time we will find
\eqn
\log g_{i\bar i} &=& \log~\prod_{j=1}^i ~\vev{\bar j|\sigma|j}^2\no\\
&\rightarrow& \int^s ds'~ \log u \\
&\sim& \int^s ds'~ H\no
\enq
by (\ref{uresult}).
But this is exactly the definition of $H$, so the picture is consistent.

\sect{Concluding Remarks}

We showed that the `new index' for the \CP\ model computed with large $N$
methods agreed with that determined by the $tt^*$ formalism.
As in previous work, the formalism seems to be a fertile source
of pretty mathematical structure.
Much more is known about the basic structure we saw here (continuous Toda
and Legendre transforms) than we made use of.\cite{Krichever,Takabe}
An analogous computation of the index for finite $N$,
or for other integrable $N=2$ theories,
can be done using the TBA, \cite[x]{cfiv}
but it seems out of reach at present to relate it analytically to the
$tt^*$ equation.
It would be very interesting to know if some version of the structure here
generalizes to finite $N$.

Equally importantly, we feel the present work is a step towards
a physical understanding of the $tt^*$ formalism.
We find the formalism attractive, not just for classifying $N=2$ theories,
but
as a prototype of an exact result, revealing dynamical information about a
full
quantum field theory, but not requiring exactly solving the full theory to
get,
which is something we would very much like to have in higher dimensions.
As an illustration, the large $N$ \CP\ model has the great advantage
that an effective action can be derived,
in terms of which every element of the
formalism can be realized classically.
We can see in what sense the formalism is a reduction to $D=1$,
and where two-dimensional physics enters.
The dynamics visible in the formalism is simple but non-trivial --
in this model, particles interact only through the constant modes of the
auxiliary fields.

Since we have an exact, non-perturbative result,
we can reevaluate the old debate\cite{wit2,affleck,jevicki,polyakov}
on the importance of instantons in the large $N$ limit.
The role of the instanton correction $x^N=\beta^N$
in the chiral ring is to produce the boundary condition $H(s)=H(s+1)$.
This boundary condition is satisfied for the true $tt^*$
solution but causes the semiclassical approximation to the
$\beta\rightarrow 0$
boundary condition to be non-analytic at $s=0$.
If we ignored this, we could define
a formal limit of the model in which (\ref{smallbeta}) determines
the metric and in which non-perturbative effects are neglected.
As in
\cite[x]{affleck} the non-perturbative effects being neglected
are solitons, which contribute $O(\exp -\beta\mu e^{-1/g^2})$.
The $tt^*$ solution in this limit is singular.
It is an intriguing aspect of the formalism that regularity of the
solutions
of the non-linear $tt^*$ equations gives strong restrictions on allowed
boundary conditions.
Here the question is whether there is a unique solution with
asymptotics described by (\ref{smallbeta})
and (\ref{largebeta}).
It would be quite interesting to make such a statement
for (\ref{oureqn}).
It seems likely that such a statement would depend on the boundary
condition
$H(\beta,s)=H(\beta,s+1)$ and
in this formal sense we would say that instanton effects
are important in the model.

It should be possible to make the spectral flow construction explicit
as well.
It would also be nice to know more about the self-dual metric $K$ and
whether it has a physical interpretation in our problem.

\vspace{0.3cm}

M.B. thanks P. Fendley and R. Sorkin for useful discussions and
acknowledges the support of a Glasstone fellowship and of
St Hilda's college.
M.R.D. thanks K. Intriligator, V. Kazakov, I. Singer and C. Vafa for
valuable discussions, and the NSF and Sloan Foundation for their support.
We also thank the Aspen Center for Physics, where part of this work
was carried out.

\goodbreak

\end{document}